\documentclass[a4paper]{jpconf}
\usepackage{graphicx}
\usepackage{graphicx}
\usepackage{amsmath}
\usepackage{dcolumn}
\usepackage{epsfig}
\usepackage{psfrag}
\usepackage{lineno}
\usepackage{color}

\RequirePackage{xspace}
\RequirePackage{ifthen}

\DeclareGraphicsExtensions{.epsi,.eps,.ps,.eps.gz,.ps.gz,.gif,.epsi.gz}
\usepackage{relsize}
\def\babar{\mbox{\slshape B\kern-0.1em{\smaller A}\kern-0.1em
    B\kern-0.1em{\smaller A\kern-0.2em R}}\xspace}

\def\belle{Belle\xspace}

\def\KS    {\ensuremath{K^0_{\scriptscriptstyle S}}\xspace}

\def\KL    {\ensuremath{K^0_{\scriptscriptstyle L}}\xspace}

\def\B       {\ensuremath{B}\xspace}
\def\Bz      {\ensuremath{B^0}\xspace}
\def\Bzb     {\ensuremath{\overline{B}^0}\xspace}

\def\CP      {\ensuremath{C\!P}\xspace}

\def\CPT     {\ensuremath{C\!PT}\xspace}
\def\C       {\ensuremath{C}\xspace}
\def\P       {\ensuremath{P}\xspace}
\def\T       {\ensuremath{T}\xspace}

\def\pep2{PEP-II}
\def\BF{$B$ Factory\xspace}
\def\BFs{$B$ Factories\xspace}

\begin{document}
\title{Symmetry violations at \babar}

\author{Adrian John Bevan}

\address{Queen Mary University of London, Mile End Road, E1 4NS, United Kingdom}

\ead{a.j.bevan@qmul.ac.uk}

\begin{abstract}
Following a brief introduction I report the current status of symmetry violation tests from the 
\babar experiment, including recent results on the measurement of \T violation, and searches
for \CP and \T violation in mixing.  
\end{abstract}

\section{Introduction}
%
%
The study of the discrete symmetries \C (charge conjugation), \P (parity), \CP, \T (motion reversal), 
and \CPT has provided many insights as to the underlying structure of the weak interaction.  
Noether's theorem reminds us that in general we expect that exact symmetries in nature relate to
conservation laws.  Hence, any conserved symmetry should lead to a corresponding conservation 
law that underpins some physical phenomenon.  Likewise symmetry violation would lead to the 
consequence that the conservation law did not work exactly, but was at best a rule of thumb that
may be used to make estimates with some degree of uncertainty regarding the accuracy of any
conclusion inferred; for example SU(3) falls into this category of symmetry.
The symmetries \C and \P are known to be broken, and the violation of parity discovered by Wu
in 1957~\cite{Wu:1957my} was the first experimental signal of how important it was to understand
violation of these symmetries for weak interactions.  Parity violation led inextricably to the 
$V-A$ structure of the weak interaction in the Standard Model of particle physics.  The fact that
there are weak decays to flavour specific final states also highlights the presence of \C violation.
However a remarkable thing at that time was that \CP was found to be conserved.  Hence the large
violation of parity and charge conjugation was such that it balanced in the combination \CP.  This
assertion remained true until 1964, when Cronin, Fitch, Christenson and Turlay discovered \CP
violation in neutral kaon decay as a consequence of trying to improve our understanding of 
regeneration~\cite{Christenson:1964fg}.  It was quickly noted by Sakharov as to how important
\CP violation was for cosmology~\cite{Sakharov:1967dj}.  Without \C and \CP violation the
matter and antimatter manifest in the big bang would annihilate each other and the residual 
Universe would have been devoid of the large imbalance that we observe today.  It is interesting
to note that since that discovery we have found several other manifestations of \CP violation,
and also tested the level of \T violation.  However it took 45 years from the initial discovery
for direct \CP violation to be found in kaons~\cite{AlaviHarati:1999xp}\cite{Fanti:1999nm}.  A few years after this, large (10\% level) 
\CP violation was found studying the longitudinal polarisation basis of a rare \KL decay to a four body final 
state~\cite{Abouzaid:2005te}\cite{Lai:2003ad}. On the theoretical side a model of \CP violation was proposed by
Kobayashi and Maskawa, building on earlier work concerning quark mixing by Cabibbo~\cite{Cabibbo:1963yz}\cite{Kobayashi:1973fv}.  With hindsight,
after several decades of detailed experimentation,
it turns out that this model is the correct leading order description of \CP violation in
the SM.  It is worth noting that just as in the case of \C and \P violations balancing to conserve
\CP (most of the time), that the measured levels of \CP and \T violation balance such that 
overall \CPT is conserved.  The ramifications of \CPT conservation include Lorentz invariance, which underpins our understanding of modern
physics.  At some energy scale it is expected that quantum effects will become important when
trying to describe space-time, and a consequence of this would be Lorentz violation, and 
hence \CPT violation.  Thus far however, experimental evidence continues to support both Lorentz and \CPT invariance. 
The remainder of these proceedings focus on the \BF tests of these
discrete symmetries from the perspective of \babar results, with a brief historical 
interlude followed by more recent results.  The avid reader should refer to the Physics of the \BFs for
more details~\cite{Bevan:2014iga}.

\section{\CP violation in \B meson decay}
%
%
The \BFs, the \babar experiment and \pep2 collider at the SLAC National Accelerator Laboratory
and the Belle experiment and KEKB collider at KEK, were built to discover \CP violation in 
the decay of neutral \B mesons.  This primary goal drove the teams building those experimental
facilities to build in safety margins in order to be sure that the mission would be 
accomplished, even if \CP violation turned out to be much smaller than expected by many theorists
of the day.  Nature was kind, and in 2001 \babar unveiled results showing \CP violation in
the decay of a $b$ quark to a $c\overline{c}s$ final state (modes such as $\B\to J/\psi \KS$, 
and $J/\psi \KL$ were included in this discovery).  That was quickly confirmed by \belle, 
firmly establishing that indeed \CP violation was exhibited in \B decay, and that it
was large.  These two results were published as back-to-back articles in 
Phys. Rev. Lett.~\cite{Aubert:2001nu}\cite{Abe:2001xe}.  In contrast with kaons, it only took 
a further two years before direct \CP violation was established in $\B$ decays via
the decay $B\to K^\pm \pi^\mp$.  Both \CP violation in the interference
between mixing and decay amplitudes and direct \CP violation turned out to be order one effects in \B decays,
in contrast to the $10^{-3} - 10^{-6}$ levels seen in kaons.  The foresight that led to
including safety margins in the design of the \BFs has resulted in a richer and broader
physics programme than originally envisaged.

\section{Other symmetry tests in \B meson decay}
%
%
Banuls and Bernabeu proposed using entangled pairs of \B mesons to test \T, \CP, and \CPT asymmetries~\cite{Banuls:2000ki}
as a generalisation of the Kabir asymmetry measurement proposed in 1970~\cite{Kabir:1970ts} that used only a flavour filter basis.  
Using two pairs of orthonormal states, one based on $b$-quark flavour and one based on \CP eigenvalue, \babar
are able to measure 12 asymmetries of which there are four distinct tests of each of \T, \CP, and \CPT~\cite{Lees:2012kn}.
The methodology relies on comparing some reference process with that of the symmetry conjugated
one as a function of proper time between two events for the entangled pair of mesons evolving into their 
flavour and \CP decay filters.  The first event is marked by the collapse of the entangled wave 
function (at some time $t_1$) which coincides with the decay of one of the $B$ mesons in the 
entangled pair, and the second event is marked by the decay of the remaining $B$ (at time $t_2$).
For example one of the four tests of \T involves comparing the time-dependent ($t_2-t_1 > 0$) rate of
$\overline{B}^0 \to B_-$ with that of the conjugate transition $B_- \to \overline{B}^0$.\footnote{Here the $-(1)$ subscript refers
to the \CP eigenvalue of the \B decay filter; in this case it is \CP odd, corresponding to $J/\psi \KS$.  The other
pairings of reference and transformed transitions can be found in Refs.~\cite{Banuls:2000ki}\cite{Lees:2012kn}.}
The standard \babar flavour tagging techniques developed for the \CP violation discovery measurement
are used in order to select \Bz and \Bzb mesons, and the selection of \CP even and \CP odd filter
basis pairs, for $b\to c\overline{c}s$ transitions differ by a \KL vs a \KS in the final state 
(see for example~\cite{Aubert:2009aw}).  As reported in \cite{Lees:2012kn}, the 
\babar data are consistent with both \CP and \T violation, whilst \CPT is conserved.  
The significance reported for the observation of \T violation (assuming Gaussian uncertainties) exceeds $14\sigma$.
Thus the level of \CP and \T violation balance each other.  A number of related measurements have
recently been proposed~\cite{Bevan:2013rpr}\cite{Dadisman:2014mya} and Ref.~\cite{Schubert:2014ska} is a recent review on \T violation
measurements in mesons, which may be of interest to the reader.

%
%
Using only a flavour filter basis pair one can define two additional asymmetries.  One of these is 
the Kabir asymmetry, which tests \CP and \T (in mixing), and the other tests \CP and \CPT.
The corresponding asymmetry is given by
\begin{eqnarray}
A_{\CP, \T} = \frac{\Gamma(\Bzb\to \Bz) - \Gamma(\Bz\to \Bzb)}{\Gamma(\Bzb\to \Bz) + \Gamma(\Bz\to \Bzb)} = \frac{1-|q/p|^4}{1+|q/p|^4} = \frac{N^{++} - N^{--}}{N^{++} + N^{--}},
\end{eqnarray}
where $q$ and $p$ are parameters related to mixing and $N^{++}$ and $N^{--}$ are 
event yields.
The SM expectation for $A_{\CP, \T}$ is $(-4.0\pm 0.6)\times 10^{-4}$~\cite{Lenz:2006hd}\cite{Charles:2011va}\cite{Lenz:2012az},
which is beyond current experimental reach, however large enhancements are possible in the 
presence of physics beyond the SM (c.f. the D0 measurement of this quantity for $\B_{s}$ mesons~\cite{Abazov:2011yk}\cite{Abazov:2012hha}).
Traditional measurements searching for \CP (and \T) violation
in mixing use semileptonic decays of \B mesons (see for example Section 17.5 of ~\cite{Bevan:2014iga}).  
The experimental signature typically used for this measurement
is a pair of semi-leptonic decays reconstructed with same sign leptons, as the first meson decays to 
fix the initial flavour of the un-decayed \B, and that meson subsequently mixes before decay. 
However those results are systematically limited by the size of control samples to estimate the level of 
wrong sign events.  As a consequence it is interesting to attempt other ways of performing the same test.
\babar have recently performed a measurement using a
combination of hadronic and semi-leptonic tags in order to explore an experimentally distinct
final state with the hope of bypassing the traditional systematic limitations. The semi-leptonic
decay $\Bz\to D^{*-}X\ell^+\nu$ with a partially reconstructed $D^{*-}\to \pi^- \overline{D}^0$ is used for this 
measurement where the asymmetry between $\ell^+$ and $\ell^-$ observed in data includes the 
effect of \B mixing, and of charge reconstruction bias in the detector, both of which have to be corrected for
in order to determine $A_{\CP, \T}$.  Hadronic \B decays used in this analysis are ``kaon tagged'' events, 
where a neutral \B meson decays into a final state with a charged kaon.  The charge asymmetry observed in
data for these events, in addition to including an overall detector bias, has a correction from the 
asymmetry in reconstructing the charged kaon, arising from the difference in nuclear cross sections
of $K^+$ and $K^-$ in detector material.
The measured value of $A_{\CP, \T}$ is $(0.06\pm 0.17^{+0.38}_{-0.32})\%$, which is compatible with 
expectations. However, the uncertainties are considerably larger than the SM level and dominated by
systematic uncertainties associated with peaking background contributions and the $\mathrm{\Delta}t$
(the proper time difference between the decay of the semi-leptonic and hadronic \B decays in the event)
resolution function model.  Hence these uncertainties can be further reduced with input from larger data
samples in the future.

\babar has also measured \CP asymmetries for a number of rare decays, and the contribution by Simon Akar
to these proceedings gives a review of some of those recent measurements.

\section{Symmetry violation searches in charm decay}
%
%
\babar has also searched for \CP violation in charm decays.  The details can be found in~\cite{Bevan:2014iga},
where time-integrated and time-dependent analyses have been performed.  In recent years there
has been a resurgence of interest in this area for two reasons (i) charm mixing is now firmly
established, which sets the scale for mixing-dependent effects to be manifest and opens the 
door for such measurements, and (ii) in 2011 LHCb reported time-integrated \CP asymmetry difference between
$D\to K^+K^-$ and $\pi^+\pi^-$ modes that was non trivial.  The latest results are compatible with
no evidence for \CP violation~\cite{hfag}. There is also interest in the study of triple product asymmetries 
that probe \CP violation using the longitudinal polarisation basis of a decay (c.f. kaon decays), in particular the
decay modes $D^0\to K^+K^-\pi^+\pi^-$, $D^+\to \KS K^+ \pi^+\pi^-$, and $D^+_s\to \KS K^+ \pi^+\pi^-$
have been studied~\cite{delAmoSanchez:2010xj}\cite{Lees:2011dx}.   The results are consistent with \CP conservation, in line with expectations
that the charm system exhibits small weak phase differences in the SM, hence small levels of \CP violation.
A recent re-analysis of the \babar data was presented at CKM `14~\cite{ckm2014} using the
methodology introduced in Ref.~\cite{Bevan:2014nva}.  This shows non-trivial \C and \P violations
for the $D^0$ and $D^+_s$ decay, but not for $D^+$.  These results may ultimately yield deeper insights on the weak
interaction and hadronisation.

\section{The search for \CP violation in $\tau$ decay}
%
%
Searches for \CP violation are not confined to quark interactions and following CLEO and \belle's lead, 
\babar has measured the direct \CP asymmetry of $\tau^+\to \KS\pi^+\nu$.
A non-trivial level of \CP violation is expected as a result of the neutral kaon in the final state, 
as well as \KS-\KL interference and regeneration effects.  The result obtained $A_{\CP} = (-0.36\pm 0.23 \pm 0.11)\%$ 
is compatible with no evidence for \CP violation, 
however this lies $2.8\sigma$ from the SM expectation of $+0.36\%$~\cite{BABAR:2011aa}.  

\section{Summary}
%
%
In summary \babar has performed many interesting tests
of discrete symmetries for $B$, $D$, and $\tau$ decays, and has observed several symmetry violations
in meson decay.  
While \babar stopped taking data several years 
ago, it is still producing results that are competitive and able to teach us 
something new about the behaviour of the weak interaction and of these 
discrete symmetries.

\end{document}